\documentclass{aa}  

\usepackage{graphicx}
\usepackage{txfonts}
\usepackage{siunitx,subfig,xspace,natbib}

\newcommand{\Msol}{M\ensuremath{_\odot}\xspace}

\newcommand{\mrm}[1]{\mathrm{#1}}

\begin{document}

   \title{Gamma-Ray Observations of Nova Sgr 2015 No. 2 with INTEGRAL}


\author{
  Thomas Siegert   		   \inst{\ref{inst:mpe},\ref{inst:xcu}}\thanks{E-mail: tsiegert@mpe.mpg.de} \and
	Alain Coc              \inst{\ref{inst:csnsm}} \and
	Laura Delgado          \inst{\ref{inst:csic},\ref{inst:ieec}} \and
  Roland Diehl           \inst{\ref{inst:mpe},\ref{inst:xcu}} \and
	Jochen Greiner         \inst{\ref{inst:mpe}} \and
  Margarita Hernanz      \inst{\ref{inst:csic},\ref{inst:ieec}} \and
  Pierre Jean            \inst{\ref{inst:irap}} \and
  Jordi Jose             \inst{\ref{inst:upc},\ref{inst:ieec}} \and	
	Paolo Molaro           \inst{\ref{inst:inaf}} \and
	Moritz M. M. Pleintinger \inst{\ref{inst:mpe}} \and
	Volodymyr Savchenko    \inst{\ref{inst:isdc}} \and
	Sumner Starrfield      \inst{\ref{inst:asu}} \and
	Vincent Tatischeff     \inst{\ref{inst:csnsm}} \and
	Christoph Weinberger   \inst{\ref{inst:mpe}}
}
\institute{
  Max-Planck-Institut f\"ur extraterrestrische Physik, Gie\ss enbachstra\ss e, D-85741 Garching, Germany
  \label{inst:mpe}
  \and
	Excellence Cluster Universe, Boltzmannstra\ss e 2, D-85748, Garching, Germany
  \label{inst:xcu}
  \and
	Centre de Sciences Nucl\'eaires et de Sciences de la Mati\`ere (CSNSM), CNRS/IN2P3, Univ. Paris-Sud, Universit\'e Paris–Saclay, F–91405 Orsay Campus, France
  \label{inst:csnsm}
	\and
	Institute of Space Science (ICE, CSIC), Campus UAB, C/ Can Magrans s/n, E-08193, Cerdanyola del Valles (Barcelona), Spain
  \label{inst:csic}
  \and
	Institut d'Estudis Espacials de Catalunya (IEEC), E-08034, Barcelona, Spain
  \label{inst:ieec}
  \and
	CNRS, IRAP, 9, avenue du Colonel Roche, BP 44346, F-31028, Toulouse Cedex 4, France
	\label{inst:irap}
	\and
  Dept. de Fisica, Universitat Politecnica de Catalunya, BarcelonaTech, EEBE, C. Eduard Maristany 10, E-08019, Barcelona, Spain
  \label{inst:upc}
	\and
  INAF, Osservatorio Astronomico di Trieste, Via G. B. Tiepolo 11, I-34131, Trieste, Italy
  \label{inst:inaf}
	\and
  ISDC, Department of astronomy, University of Geneva, chemin d’Ecogia, 16 CH-1290, Versoix, Switzerland
  \label{inst:isdc}
	\and
  School of Earth and Space Exploration, Arizona State University, Tempe, Arizona 85287 1404, USA
	\label{inst:asu}	
  }

   \date{Received 21 Dec 2017; accepted 15 Mar 2018}

 
  \abstract
   {INTEGRAL observed Nova Sgr 2015 No. 2 (V5668 Sgr) around the time of its optical emission maximum on March 21, 2015. Studies at UV wavelengths showed spectral lines of freshly produced $^7\mathrm{Be}$. This could be measurable also in gamma-rays at 478~keV from the decay to $^7\mathrm{Li}$. Novae are also expected to synthesise $^{22}\mathrm{Na}$ which decays to $^{22}\mathrm{Ne}$, emitting a 1275~keV photon. About one week before the optical maximum, a strong gamma-ray flash on time-scales of hours is expected from short-lived radioactive nuclei, such as $^{13}\mrm{N}$ and $^{18}\mrm{F}$. These nuclei are $\beta^+$-unstable, and should yield emission up to 511~keV, but which has never been observed from any nova.}
   {The spectrometer SPI aboard INTEGRAL pointed towards V5668 Sgr by chance. We use these observations to search for possible gamma-ray emission of decaying $^7\mathrm{Be}$, and to directly measure the synthesised mass during explosive burning. We also aim to constrain possible burst-like emission days to weeks before the optical maximum using the SPI anticoincidence shield (ACS), i.e. at times when SPI was not pointing to the source.}
   {We extract spectral and temporal information to determine the fluxes of gamma-ray lines at 478~keV, 511~keV, and 1275~keV. Using distance and radioactive decay, a measured flux converts into the $^7\mathrm{Be}$ amount produced in the nova. The SPI-ACS rates are analysed for burst-like emission using a nova model light-curve. For the obtained nova flash candidate events, we discuss possible origins using directional, spectral, and temporal information.}
   {No significant excess for the 478~keV, the 511~keV, or the 1275~keV lines is found. Our upper limits ($3\sigma$) on the synthesised $^7\mathrm{Be}$ and $^{22}\mathrm{Na}$ mass depend on the uncertainties of the distance to V5668 Sgr: The $^7\mathrm{Be}$ mass is constrained to less than $4.8 \times 10^{-9}\,(d/\mrm{kpc})^2\,\mrm{M_{\odot}}$, and the $^{22}\mathrm{Na}$ mass to less than $2.4 \times 10^{-8}\,(d/\mrm{kpc})^2\,\mrm{M_{\odot}}$. For the $^7\mathrm{Be}$ mass estimate from UV studies, the distance to V5668 Sgr must be larger than 1.2~kpc ($3\sigma$). During three weeks before the optical maximum, we find 23 burst-like events in the ACS rate, of which six could possibly be associated with V5668.}
   {}


   \maketitle
%

\section{Introduction}

\subsection{Nova Sagittarii 2015 No. 2 / V5668 Sgr}

On 15 March 2015, Nova Sagittarii 2015 No. 2 (V5668 Sgr, short V5668) was detected by \citet[][galactic coordinates $(l_0/b_0) = (5.38^{\circ}/-9.87^{\circ})$]{Seach2015_V5668}. After a six-day rise in brightness, V5668 reached its optical maximum on March 21.67 UT, corresponding to $T_0 = \mathrm{MJD}\,57102.67$, with a V-band magnitude of 4.32~mag. Two independent studies \citep{Molaro2016_V5668,Tajitsu2016_V5668} measured blue-shifted spectral lines of singly ionised Be II at wavelengths around 313~nm. The Doppler-velocities of these UV line profiles range between -700 and $-2200~\mrm{km~s^{-1}}$. Based on a canonical ejected mass of $10^{-5}~\mrm{M_{\odot}}$ for novae, the measured abundance ratios allowed to estimate the mass of synthesised and ejected $^7\mathrm{Be}$ for V5668 from these UV measurements of $7\times10^{-9}~\mrm{M_{\odot}}$ \citep{Molaro2016_V5668}. Novae have only recently been verified as significant sources of $^7\mathrm{Li}$ in the Galaxy by detections of $^7\mathrm{Li}$ I at $6708~\mrm{\AA{}}$ in Nova Centauri 2013 \citep[V1369 Cen,][]{Izzo2015_novaeLi}, and the $^7\mathrm{Be}$ II doublet at $313.0583~\mrm{nm}$ and $313.1228~\mrm{nm}$ in Nova Delphini 2013 \citep[V339 Del,][]{Tajitsu2015_novaBe7}. 

The distance to V5668 is not precisely known. Based on expansion velocity measurements and geometrical considerations, \citet{Banerjee2016_v5668} estimated a distance\footnote{\citet{Banerjee2016_v5668} provide three distance estimates, one considering the geometry and expansion velocity, but without uncertainties, and two others, based on two different MMRD method assumptions, leading to two different absolute magnitudes, $M_V=-6.91\pm0.40$ and $-6.65\pm1.82$, and thus two distances of 1.31-1.76 and 0.68-3.6~kpc, respectively.} of $1.54~\mrm{kpc}$. \citet{Jack2017_V5668} derived a similar estimate of 1.6~kpc, using a distance modulus approach, but without providing uncertainties. The "maximum magnitude vs. rate decline" (MMRD) method to determine the distance to novae, as reviewed by \citet{dellaValle1995_nova} \citep[see also][]{Schmidt1957_novae2}, provides similar distance estimates but with an intrinsically large uncertainty. Due to the fact that the luminosity of a nova outburst depends on more physical parameters than only the white dwarf mass \citep{dellaValle1995_nova}, the distance to a single nova, as opposed to a population, might be uncertain by $\approx50\%$.  

V5668 Sgr has been observed at many wavelengths. In mid- \citep[e.g.][with SOFIA/FORCAST]{Gehrz2015_v5668SOFIA} and near-infrared \citep[e.g.][using NICS]{Banerjee2016_v5668} observations, clear signatures of dust have been seen, and also a strong detection of CO in emission \citep{Banerjee2015_v5668CO}. The total dust mass produced by V5668 is estimated to be about $2.7\times10^{-7}~\mrm{\Msol}$ \citep{Banerjee2016_v5668}, so that the mass of the gaseous component of the ejecta is between $2.7$ and $5.4\times10^{-5}~\mrm{\Msol}$. Note that the ejecta mass is difficult to estimate, because the absolute flux measurements have to be scaled by the distance to a nova, which is model-dependent and often uncertain by several tens of per cent, and also because there is no adequate observable to estimate the accretion rate. In the optical, V5668 was monitored for more than 200 days after the outburst \citep[][using TIGRE]{Jack2017_V5668}, finding transient Balmer and Paschen lines of H, different Fe II lines, as well as N I and N II. In general, the spectral shapes change during the evolution of the optical light curve, whereas after the deep minimum at day 110 after the optical maximum, clear double-peak profiles are observed. The spread in Doppler-velocities (expanding shell velocity) is up to $2000~\mrm{km~s^{-1}}$. About 95 days after the outburst, the nova was detected in soft X-rays \citep[][with Swift/XRT]{Page2015_V5668Xrays_a}, softening and brightening up to $(6.0\pm0.5)\times10^{-2}~\mrm{cts~s^{-1}}$ until day 161. During this time, the apparent H-column density towards V5668 decreased from $N_H = (4.7^{+1.6}_{-1.2})\times10^{22}~\mrm{cm^{-1}}$ to $(0.4\pm0.1)\times10^{22}~\mrm{cm^{-1}}$, while the plasma temperature increased from $1.3^{+0.5}_{-0.3}~\mrm{keV}$ to $3.4^{+1.4}_{-0.8}~\mrm{keV}$. This spectral change may be connected to the destruction of dust by soft X-ray and UV emission \citep[][Swift/UVOT]{Page2015_V5668Xrays_a}.

\begin{figure}[!ht]%
\centering
\includegraphics[width=\columnwidth,trim=1.6cm 1.6cm 3.0cm 2.7cm,clip=true]{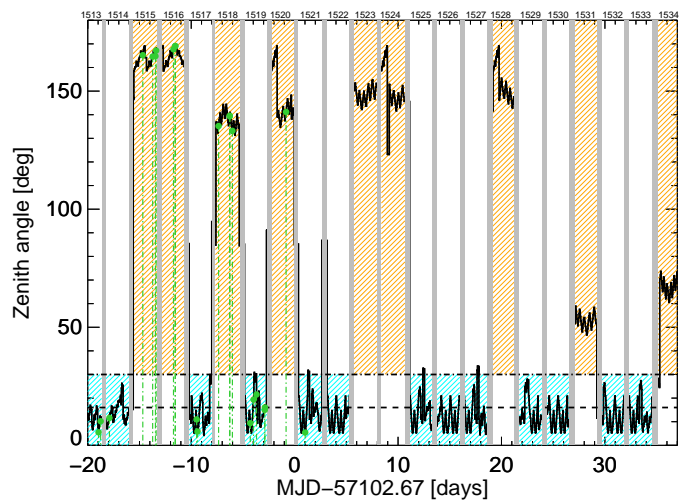}%
\caption{Position of V5668 Sgr (zenith-angle, black data points) with respect to the SPI on-axis frame as a function of time in days from the nova optical maximum (MJD 57102.67). INTEGRAL revolution numbers are labelled on the top. The black dashed and dot-dashed horizontal lines correspond to the SPI fully-coded ($16^{\circ}$) and partially-coded ($30^{\circ}$) field of view, respectively. The hatched cyan areas mark the times used for SPI on-axis analysis, whereas during the times marked by the orange hatched areas, off-axis analysis was performed using the ACS (see Sec.~\ref{sec:integralresponse}). The grey regions show the perigee-passages of INTEGRAL, i.e. when no data is taken. The green dots indicate the times when the SPI-ACS search for flash-like events found its candidates (see Sec.~\ref{sec:acsanalysis}; note that all aspect angles have been used for this search, but only during days -20 to days +3).}%
\label{fig:novadirection}%
\end{figure}

High-energy gamma-rays have also been observed for V5668 \citep[0.1-100~GeV][using Fermi/LAT]{Cheung2016_V5668Fermi}, beginning about two days after the optical maximum, as measured from similar nova outbursts \citep[e.g.][]{Hays2013_novadelFermi,Cheung2013_novacenFermi,Ackermann2014_novaeFermi}. In the $>100$~MeV band, V5668 was visible for 55 days with an average flux of $\approx 10^{-7}~\mrm{ph~cm^{-2}~s^{-1}}$, fainter than observed for this type of source \citep[][note that less than ten gamma-ray novae have been detected until the write-up of this paper]{Ackermann2014_novaeFermi}. Even though the gamma-ray emission appears sporadic and maybe delayed to the optical emission, it seems that the hard gamma-ray flux correlates with the optical light curve. In fact, for the brightest gamma-ray nova detected so far, ASASSN-16ma \citep{Luckas2016_ASASSN-16ma}, the optical light-curve strongly correlates with the 0.1-300~GeV flux during the decline phase. The gamma-ray-to-optical-flux ratio remained constant at a value of $\approx 0.002$. This tight correlation made the authors to cast doubt on the standard model for optical nova emission. Nuclear burning on the surface of the white dwarf (see Sec.~\ref{sec:explosivenova}) results into freshly-synthesised nuclei which are ejected, decay and/or de-excite, and emit MeV gamma-rays. Some of these gamma-rays may undergo Compton scattering, which powers a continuum with a low-energy cut-off around 20-30~keV, due to photo-electric absorption \citep{Gomez-Gomar1998_novae}. The optical emission, on the other hand, is mainly thermal radiation from the heated, now cooling because expanding, gas. Even though the nova explosion is triggered by nucleosynthesis reactions, the resulting $\sim\,\mrm{MeV}$ gamma-rays provide only a small amount of energy during the envelope expansion. The visual maximum corresponds therefore to the maximum expansion of the photosphere. After the optical peak, the photosphere recedes, and higher temperatures become visible, so that the peak moves to UV wavelengths. But this scenario cannot explain the high-energy GeV gamma-ray emission, which is generally attributed to shock-accelerated particles, and even more so cannot explain the correlation between the optical and GeV emission. Instead, the authors propose that the optical emission is predominantly originating also in the shocks, rather than in the photosphere above. This automatically explains the simultaneous emission and also the questionable super-Eddington luminosities observed for many novae, because shocks may not be treated as hydrostatic atmospheres \citep[see also][]{Martin2017_novagammas}. In the case of V5668, the Fermi/LAT $>100$~MeV light-curve also seems to show multiple peaks, so that this nova might have had multiple mass ejections, and thus there might be multiple onsets of nucleosynthesis. If a local maximum in the optical or $>100$~MeV light-curve indeed corresponds to an additional mass ejection, the estimates of this and other works might be more uncertain, as each individual outburst might eject its own amount of mass. In this paper, we focus on a single time origin of explosive burning, and discuss only one major mass ejection.

This shock scenario also predicts hard X-ray emission at later times, contemporaneous to the GeV emission, depending on the nova outflow properties, such as density, mass, velocity, and the resulting optical depth. The peak of such an additional X-ray emission would be expected between 30 and 210 days after the optical maximum \citep{Metzger2014_novashocks}. These non-thermal X-rays would be produced by optical (eV) photons, being Compton up-scattered on GeV particles. Following \citet{Metzger2015_novashocks}, an order of magnitude estimate for the expected X-ray flux can be derived from the measured GeV flux $F_{GeV}$. This assumes a fraction $f_X$ of high-energy gamma-ray luminosity to be radiated away in X-rays of energy $E_X$, so that the resulting X-ray flux is $F_X \sim f_X F_{GeV} / E_X$. For example, an $f_X$-value of 0.01 would predict a hard X-ray flux of the order of $10^{-6}~\mrm{ph~cm^{-2}~s^{-1}~keV^{-1}}$ for V5668 Sgr at about 50~keV. The influence of such a plausible additional flux at hard X- and also soft gamma-rays on the INTEGRAL measurements will be discussed in Sec.~\ref{sec:nucsysejecta}.

\subsection{Explosive nucleosynthesis in novae}\label{sec:explosivenova}

The nova explosion is typically explained by a thermonuclear runaway on the surface of a white dwarf. At a mass accretion rate of $10^{-10}$-$10^{-9}~\mrm{\Msol~yr^{-1}}$, the accreted matter reaches degeneracy by the strong gravitational field of the white dwarf. Once the ignition conditions for hydrogen burning are met, nucleosynthesis starts. Even though the envelope is initially degenerate, once the temperature in the envelope exceeds $3\times10^{7}~\mrm{K}$, degeneracy is lifted in the whole envelope \citep{Jose2016_stellarexplosions}. However, it is important to stress that a nova outburst likely occurs because of a hydrogen thin shell instability \citep{Schwarzschild1965_thermonuclear_stability,Yoon2004_thermonuclear_stability}, for which degeneracy is not required at all. In general, nuclei up to $A \approx 40$ are produced and ejected in a classical nova explosion. In this scenario, hydrogen burning proceeds through the CNO-cycle, i.e. it is required that such seed nuclei are present. During the CNO burning, short-lived nuclei (e.g. $^{13}\mrm{N}$, $^{14}\mrm{O}$, $^{15}\mrm{O}$, or $^{17}\mrm{F}$ with half-life times of 597, 71, 122, and 65~s, respectively) are produced, which undergo $\beta^+$-decay, and work as an energy source for the expansion of the outer, low-density shell \citep[e.g.][]{Starrfield1972_novaCNO,Jose2006_novae}. For gamma-ray observations, mainly two species are important as they emit the strongest mono-energetic gamma-ray lines, $^7\mathrm{Be}$ and $^{22}\mathrm{Na}$. In addition, $^{13}\mrm{N}$ and $^{18}\mrm{F}$ ,from the family of short-lived $\beta^+$-unstable nuclei, may be observable due to positron annihilation as a 511~keV gamma-ray flash (see below).

\begin{figure}[!ht]
	\centering
	  \subfloat[$^7\mrm{Be}$ 478~keV line region. \label{fig:spec_478}]{\includegraphics[width=0.45\textwidth,trim=0.4cm 1.6cm 0.2cm 3.1cm,clip=true]{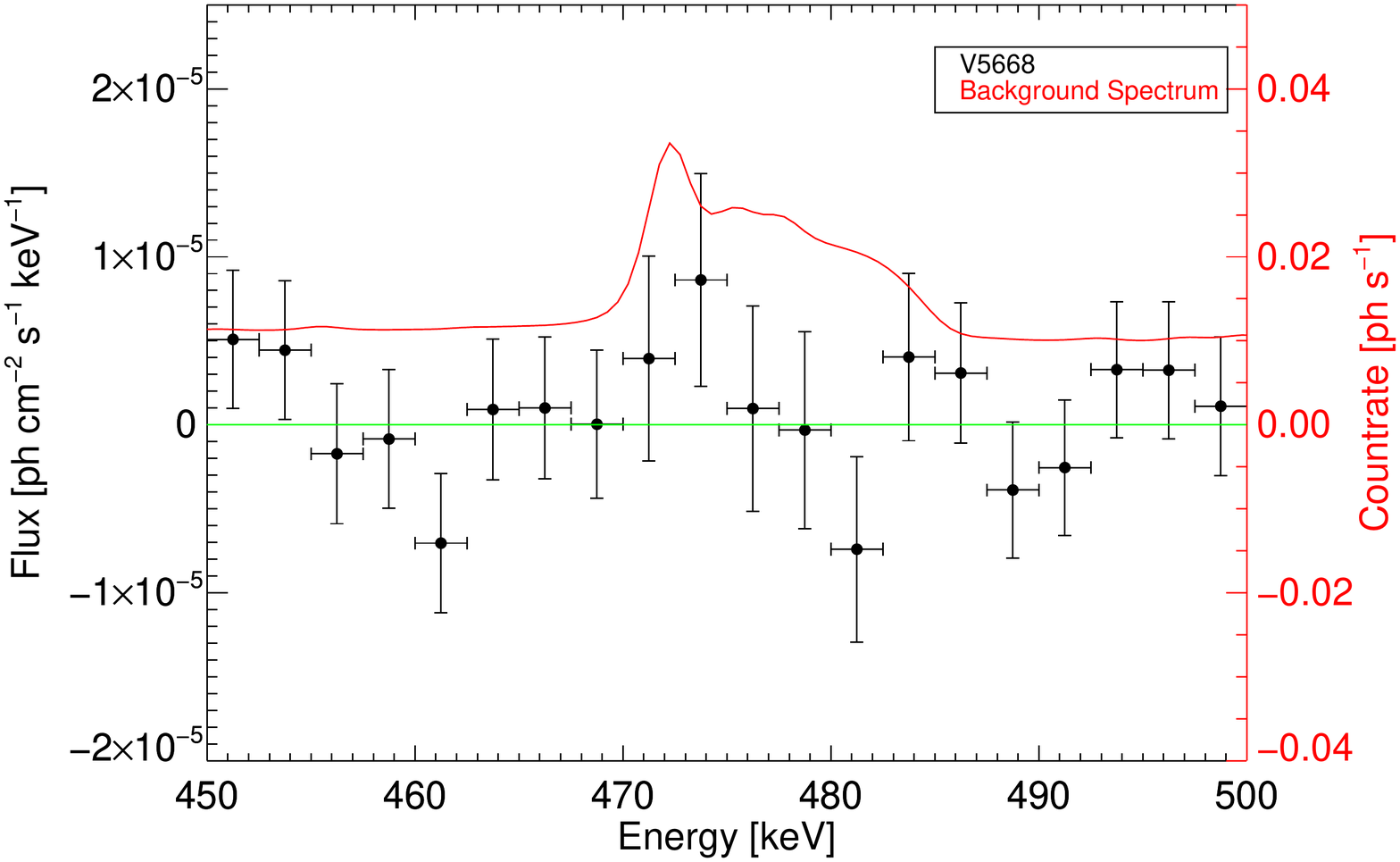}}\\
	  \subfloat[$^{22}\mrm{Na}$ 1275~keV line region. \label{fig:spec_1275}]{\includegraphics[width=0.45\textwidth,trim=0.4cm 1.6cm 0.0cm 3.1cm,clip=true]{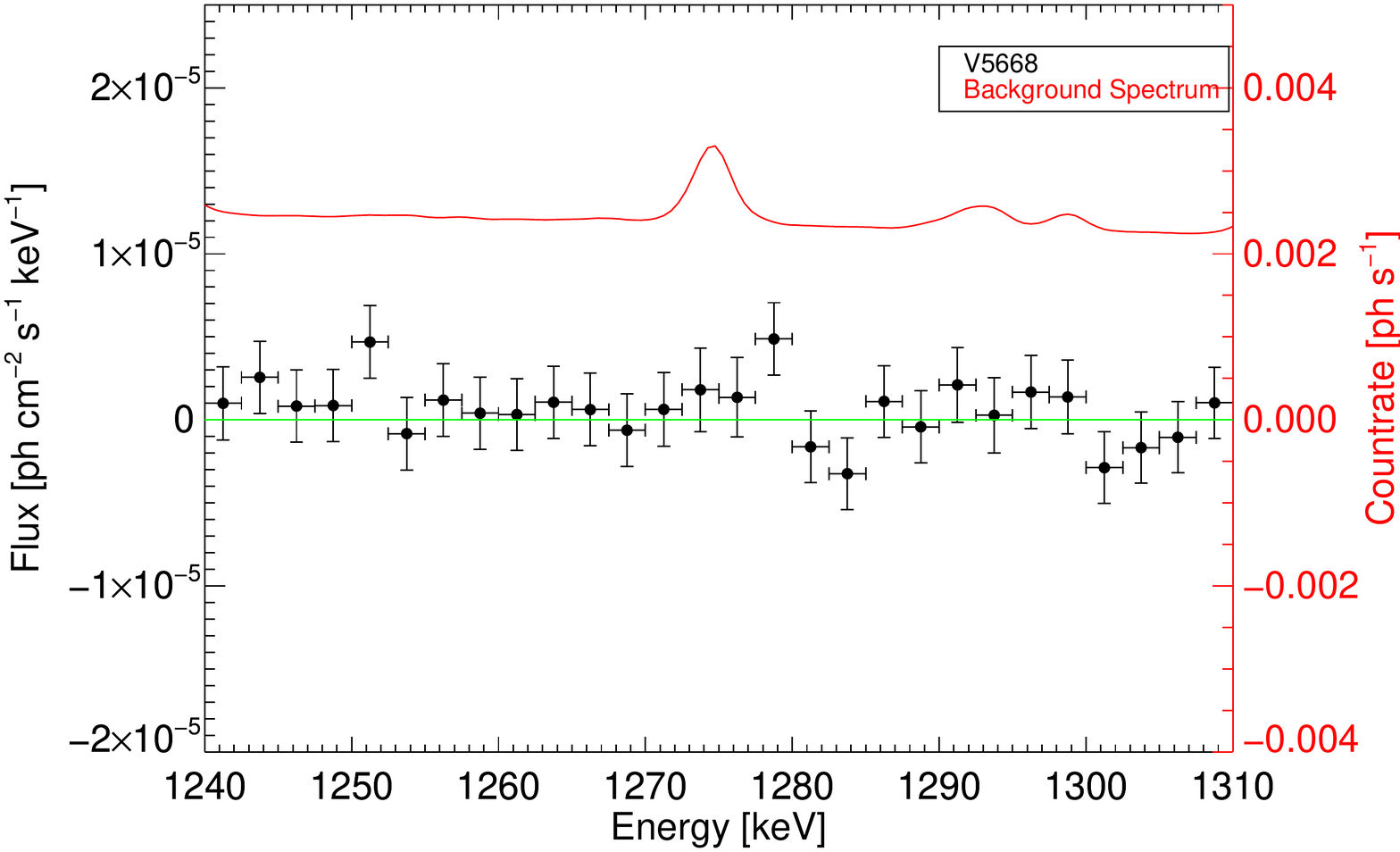}}
	  \caption{Spectra (black data points, left-hand axes, 2.5~keV binning) for V5668, integrated between days -7 and +37 after the optical maximum. Also shown are the SPI background spectra (red, right-hand axes) in the energy range 450 to 500~keV (a), and 1240 to 1310~keV (b). The background is well suppressed, and no significant line-like or continuum excesses above background can be identified. The zero-flux level is indicated by green solid lines. See text for details.}
		\label{fig:V5668_specs}
\end{figure}

The isotope $^7\mathrm{Be}$ is thought to be produced in such a nova explosion via the reaction $^3\mathrm{He}(\alpha,\gamma)^7\mathrm{Be}$. In accelerator experiments, $^7\mathrm{Be}$ is produced predominantly in the ground state, $\mrm{1/2^-}$. However, about 40\% of the time, it is created in its first excited nuclear state, $\mrm{3/2^-}$, at 429~keV \citep{Parker1963_7Be,diLeva2009_429keV}, for which reason also a 429~keV line could be expected, if the nova envelope was not opaque at this time\footnote{This is true for many other prompt gamma-ray lines which are produced during explosive burning. Here, the 429~keV line serves as a proxy for similar $(p,\gamma)$-reactions of the CNO-cycle, operating in shells of temperatures of $\approx 10^8$~K, which are expected to be opaque at this time.}. $^7\mathrm{Be}$ decays with a half-life time of $T_{1/2}^{7\mathrm{Be}} = 53.12~\mrm{d}$ (characteristic time of $\tau^{7\mrm{Be}} = T_{1/2}^{7\mathrm{Be}} / \ln(2) = 76.64~\mathrm{d}$) via electron capture to $^7\mathrm{Li}$. This daughter nucleus de-excites to its ground state after 73~fs via the emission of a gamma-ray at $E_0^{7\mrm{Be}} = 477.60~\mrm{keV}$ with a branching ratio of $p^{7\mrm{Be}} = 10.52\%$ \citep{Firestone03}. Depending on the nova model assumptions, the yield of $^7\mrm{Be}$ in CO novae ranges between $10^{-11}$ to several $10^{-9}~\mrm{M_{\odot}}$ \citep[e.g.][]{Hernanz1996_novae7Li}. In general, there are two main types of classical novae, CO- and ONe-types. On the one hand, this classification is based on the final composition of the white dwarf, reflected by the mass and hence the expected burning stages that the progenitor star underwent. Here, a CO white dwarf comes from a star only having burnt H and He, whereas an ONe white dwarf progenitor also started C burning. In return, it is also a plausible assumption, for example in nova model simulations, that white dwarfs below $\sim 1.1~\mrm{\Msol}$ are CO-rich, while more massive ones are of ONe type. On the other hand, it is based on observational properties of novae, pointing to exactly this white-dwarf-composition, for example by measuring emission or absorption lines (see above). In a CO-type nova, the peak temperature allows nuclear burning up to oxygen, with only traces of heavier nuclei. On the other hand, because ONe white dwarfs have heavier seed nuclei, such as $^{20}\mrm{Ne}$ or $^{24}\mrm{Mg}$, in their chemical composition, ONe novae may reach temperatures to also produce silicon or argon. In the latter case, large amounts of $^{22}\mrm{Ne}$ are expected to be produced via the reaction chains $^{20}\mathrm{Ne}(p,\gamma)^{21}\mathrm{Na}(p,\gamma)^{22}\mathrm{Mg}(\beta^+)^{22}\mathrm{Na}$ or $^{20}\mathrm{Ne}(p,\gamma)^{21}\mathrm{Na}(\beta^+)^{21}\mathrm{Ne}(p,\gamma)^{22}\mathrm{Na}$, and the sub-sequent decay ($T_{1/2}^{22\mathrm{Na}} = 2.6~\mathrm{yr}$) to an excited state of $^{22}\mrm{Ne}$. $^{22}\mrm{Ne}$ then de-excites by the emission of an $E_0^{22\mrm{Na}} = 1274.53~\mrm{keV}$ gamma-ray for $p^{22\mrm{Na}} = 99.96\%$ of the time \citep{Firestone03}. Although this is not expected for CO novae \citep{Hernanz2014_nova}, we perform a search for $^{22}\mrm{Na}$ in V5668 (see Sec.~\ref{sec:spianalysis}).

Theoretical studies \citep[e.g.][]{Clayton1974_novae,Leising1987_novae511,Jose2001_novaegamma,Jose2003_novae,Jose2006_novae,Hernanz2006_novae,Hernanz2014_nova} predict a gamma-ray flash around one week before the optical maximum due to short-lived isotopes, such as $^{13}\mrm{N}$ and $^{18}\mrm{F}$, which decay via positron emission. The true time lag between the initial explosive burning, which results in $\lesssim\mrm{MeV}$ gamma-ray emission, and the optical maximum is fundamentally unknown. Initially, when the nova envelope starts to expand, it is optically thick, so that low-energy photons could not escape. Depending on the nova model set-up, the lag is determined by the time of the maximum temperature (as provided by theory), and the largest extent of the photosphere (optical maximum), resulting in a temporal offset between a few days a two weeks. Throughout this paper, we use a canonical value of the onset of explosive burning of $T_0 - 7~\mrm{d} = \mathrm{MJD}\,57095.67$ when estimating line fluxes, and release this constraint when searching for the gamma-ray flash (cf. Secs.~\ref{sec:spianalysis} and \ref{sec:acsanalysis}). The produced positrons may annihilate quickly and may produce a strong gamma-ray line at 511~keV, and a hard X-ray / soft gamma-ray continuum up to 511~keV. For a nova distance of 1~kpc, the peak flux in the 75-511~keV band, including the annihilation line, may be as high as $0.2~\mrm{ph~cm^{-2}~s^{-1}}$ \citep[e.g.][]{Hernanz2014_nova}. Thus, independent of the direction of a nova with respect to INTEGRAL, this may be seen up to distances of several kpc \citep{Jean1999_novasearch}. After the peak, the flux declines sharply, and the flash may only be seen by chance.

\subsection{INTEGRAL/SPI observations of V5668 Sgr}

ESA's gamma-ray observing satellite INTEGRAL \citep{Winkler2003_INTEGRAL} was pointed to the galactic centre before and after the optical maximum of V5668. Due to the large field of view of the spectrometer SPI \citep{Vedrenne2003_SPI} aboard INTEGRAL (field of view: $16^{\circ} \times 16^{\circ}$; angular resolution: $2.7^{\circ}$), the nova was also observed, as part of other regular observations. SPI measures X- and gamma-ray photons in the energy range between 20 and 8000~keV, using high-purity Ge detectors. The spectral resolution at 478~keV is 2.1~keV (FWHM). INTEGRAL revolutions 1514, 1517, and 1519 correspond to days -18 to -16, -11 to -9, and -5 to -3 with respect to the optical maximum of V5668. After the optical maximum, data from INTEGRAL revolutions 1521 to 1534, with several observation gaps, allow for high spectroscopic resolution gamma-ray measurements with SPI until day 37, see Fig.~\ref{fig:novadirection}. The intermediate gaps can be studied in their global gamma-ray emission by the anticoincidence shield (ACS) of SPI, made of 91 scintillating BGO crystals, sensitive to photon (and particle) energies above $\approx 75~\mrm{keV}$, but without spectral information. The ACS has an almost omni-directional field of view with a rather modest angular resolution ($\approx 60^{\circ}$) and high timing capabilities ($\Delta T = 50$~ms). This allows to search for flash-like events during the expected periods around day 7 before the optical maximum. The use of the ACS to detect hard X-ray / soft gamma-ray emission from novae has been investigated by \citet{Jean1999_novasearch}.

In this paper, we report a search for gamma-ray line emission from nucleosynthesis ejecta of V5668 using INTEGRAL/SPI, as well as a search for burst-like gamma-ray emission from short-lived nuclei during explosive burning weeks before the optical maximum. In Sec.~\ref{sec:dataanalysis}, we describe the analyses of SPI and its ACS data, exploiting spectral, temporal, and directional information, and derive physical parameters. The implications of the analysis results are discussed in Sec.~\ref{sec:conclusion}.

\section{Data analysis}\label{sec:dataanalysis}

\subsection{SPI analysis}\label{sec:spianalysis}

We analyse the spectral and the temporal domain for the expected gamma-ray lines at 429, 478, 511, and 1275~keV. Starting at $T_1 = T_0 - 7~\mrm{d} = \mathrm{MJD}\,57095.67$, the total exposure in our SPI data set until day +37 ($T_2 = T_0 +37~\mrm{d} = \mathrm{MJD}\,57139.67$) is $T_{exp} = 1.53~\mrm{Ms} < \Delta T = T_2 - T_1 = 44~\mrm{d} \approx 3.80~\mrm{Ms}$. This is reduced further by observational gaps in the regular INTEGRAL observation program, data selection criteria, such as orbit phases ($0.1$-$0.9$ to avoid the Earth's radiation belts) and onboard radiation monitor rates acceptance windows (to avoid charged particle showers and solar flares), and detector dead time. The total observation time is $T_{obs} = 1.06~\mrm{Ms}$. SPI data are dominated by background from cosmic-ray bombardment of satellite and instruments, leading to decay and de-excitation photons.

The spectrometer data are analysed by a maximum likelihood method, comparing measured Ge detector data to models of celestial emission and background. We use a self-consistent, high spectral resolution, background modelling procedure \citep[e.g.][]{Diehl2014_SN2014J_Ni,Siegert2016_511,Siegert2017_PhD} to extract spectra in the energy ranges 70 to 530~keV, and 1240 to 1310~keV. In general, the modelled time-patterns (count sequences in different detectors per unit time) for each model component are fitted to the measured time-pattern of the data by minimising the Cash-statistic \citep{Cash1979_cstat}

\begin{equation}
C(D|\theta_i) = 2 \sum_k \left[ m_k - d_k \ln m_k \right]\mrm{,}
\label{eq:cstat}
\end{equation}

which accounts for Poisson-distributed photon count statistics. In Eq.~(\ref{eq:cstat}), $d_k$ are the measured and $m_k$ the modelled data, which are matched (fitted) to $d_k$ by adjusting intensity scaling parameters $\theta_{(t)}$ of the model components, which are possibly time-dependent, 

\begin{equation}
m_k = \sum_{t_S} \sum_j R_{jk} \sum_{i=1}^{N_S} \theta_{i,t_S} M _{ij} + \sum_{t_B} \sum_{i = N_S+1}^{N_S + N_B} \theta_{i,t_B} B_{ik}\mrm{.}
\label{eq:modeldesc}
\end{equation}

Here, we describe the model in each half-keV energy bin $k$ as a superposition of $N_S$ celestial models $M_{ij}$, to which the instrumental imaging response (coded-mask response, $R_{jk}$) is applied for each image element $j$, and $N_B$ background models $B_{ik}$. For the background model, we use the information gathered over the INTEGRAL mission years, separating long-term stable or smoothly varying properties, such as detector degradation (linear within half a year) or solar activity (solar cycle anti-correlates with cosmic-ray intensity rate), from short-term variations, such as solar flares and general pointing-to-pointing variations. Instrumental gamma-ray continuum and line backgrounds are treated separately, according to their different physical origins inside the satellite. Each instrumental line imprints a certain pattern onto the gamma-ray detector array, depending on the distribution of the radiating material inside the satellite. These patterns are constant over time, as the material distribution does not change. Only detector failures lead to a change in those patterns, as double scattering photons in dead detectors are then seen as single events in working neighbouring detectors. Different activation rates (cosmic-ray bombardment) and isotope decay times then lead to different amplitudes in those patterns, which are determined using Eqs.~(\ref{eq:cstat}) and (\ref{eq:modeldesc}). For any specific process, these patterns are constant in time; however, for single energy bins (typically 0.5~keV, cf. instrumental resolution of 2-3~keV), these patterns may change due to different degradation strengths in the 19 detectors. Hence, we determine the detector patterns by performing a spectral decomposition (statistical fit) in each of the detectors on a three-day (viz. one INTEGRAL orbit) time scale. This allows to trace the degradation and the general response properties of all detectors with time, and at the same time smears out celestial contributions, because the varying time-patterns of the coded-mask response in combination with the INTEGRAL dithering strategy average out. The procedure and functions to determine the spectral response parameters have already been discussed in \citet{Siegert2016_511} and \citet{Siegert2017_PhD}; see also \citet{Diehl2017_SPI} for an analysis of the 15-year SPI data base.

By performing this maximum likelihood estimation for each of the spectral bins in the energy region of interest, we create spectra for each source or general emission morphology. In the case of V5668, the only ($N_S=1$) celestial model is a point-source at the position of the nova, $M_{1j} = \theta_{1,t_S} \delta(l-l_0) \delta(b-b_0)$. The decay time of $^7\mrm{Be}$ of 77 days, and the possible gamma-ray flash before the optical maximum leads to two analysis cases: (1) Integration over the entire exposure time to obtain a maximum of sensitivity for the longer-lived nucleosynthesis products in fine energy resolution (no time-dependence, $\theta_{1,t_S} \rightarrow \theta_{1}$), and (2) several time intervals of 2-3 hours to study the transient behaviour in broader energy binning to enhance the sensitivity of observing a gamma-ray flash, and also to trace the radioactive decay of $^7\mrm{Be}$ (light-curve).	

In Figs.~\ref{fig:spec_478} and \ref{fig:spec_1275}, the average spectra for V5668 between days -7 and 37 after the optical maximum are shown. In both energy bands, no significant excess is seen, and the spectra are consistent with zero flux. We derive upper limits on the fluxes by assuming an average line-shift of $-1000~\mrm{km~s^{-1}}$ \citep{Molaro2016_V5668,Tajitsu2016_V5668}, which corresponds to shifted gamma-ray lines centroids of 479.1~keV and 1278.8~keV, respectively. The line broadening is adopted as 8~keV for the 478~keV line, and 21~keV for the 1275~keV line \citep[FWHM, e.g.][instrumental resolution 2.08~keV at 478~keV, and 2.69~keV at 1275~keV]{Hernanz2014_nova,Siegert2017_PhD}. At days $T_0+(15\pm22)$, the $3\sigma$ upper limit on the 478~keV line flux is estimated to be $8.2 \times 10^{-5}~\mrm{ph~cm^{-2}~s^{-1}}$. The $3\sigma$ upper limit on the 1275~keV line flux is $7.6 \times 10^{-5}~\mrm{ph~cm^{-2}~s^{-1}}$.

In order to derive an upper limit on the mass of $^7\mrm{Be}$, the expected flux of the 478~keV line from the radioactive decay law, 

\begin{equation}
F^{7\mrm{Be}}(t) = \frac{M^{7\mrm{Be}} p^{7\mrm{Be}}}{4\pi d^2 N^{7\mrm{Be}} u \tau^{7\mrm{Be}}} \exp\left(-\frac{t - \Delta t}{\tau^{7\mrm{Be}}}\right)\mrm{,}
\label{eq:radiodecaylaw}
\end{equation}

is used. In Eq.~(\ref{eq:radiodecaylaw}), $M^{7\mrm{Be}}$ is the synthesised mass of $^7\mrm{Be}$ seen to decay, $d$ is the distance to V5668, $N^{7\mrm{Be}} = 7$ is the number of nucleons in $^7\mrm{Be}$, $u = 1.66 \times 10^{-27}~\mrm{kg}$ is the atomic mass unit, $\Delta t$ is fixed to $7$ days before the optical maximum, and $p^{7\mrm{Be}} = 10.52\%$ is the probability of emitting a 478~keV photon after the decay. The flux limits convert to a $^7\mrm{Be}$ mass limit of $M^{7\mrm{Be}}_{3\sigma} < 4.8 \times 10^{-9}\,(d/\mrm{kpc})^2~\mrm{M_{\odot}}$, and a $^{22}\mrm{Na}$ mass limit of $M^{22\mrm{Na}}_{3\sigma} < 2.4 \times 10^{-8}\,(d/\mrm{kpc})^2~\mrm{M_{\odot}}$. Using the distance values from \citet{Banerjee2016_v5668} or \citet{Jack2017_V5668} of $\approx 1.6$~kpc, the limits on the ejected masses yield $M^{7\mrm{Be}}_{3\sigma} < 1.2 \times 10^{-8}~\mrm{M_{\odot}}$, and $M^{22\mrm{Na}}_{3\sigma} < 6.1 \times 10^{-8}~\mrm{M_{\odot}}$, respectively. \citet{Molaro2016_V5668} estimated a $^7\mrm{Be}$ mass of $7\times10^{-9}~\mrm{M_{\odot}}$ from their UV spectra. This is consistent with our limit. Assuming their amount of ejected mass, the non-detection by INTEGRAL/SPI requires the distance to the nova V5668 to be larger than 1.2~kpc ($3\sigma$ lower limit).

\begin{figure}[!ht]
	\centering
		\subfloat[429~keV line. \label{fig:lc_429}]{\includegraphics[width=0.363\textwidth,trim=1.4cm 1.7cm 2.3cm 2.3cm,clip=true]{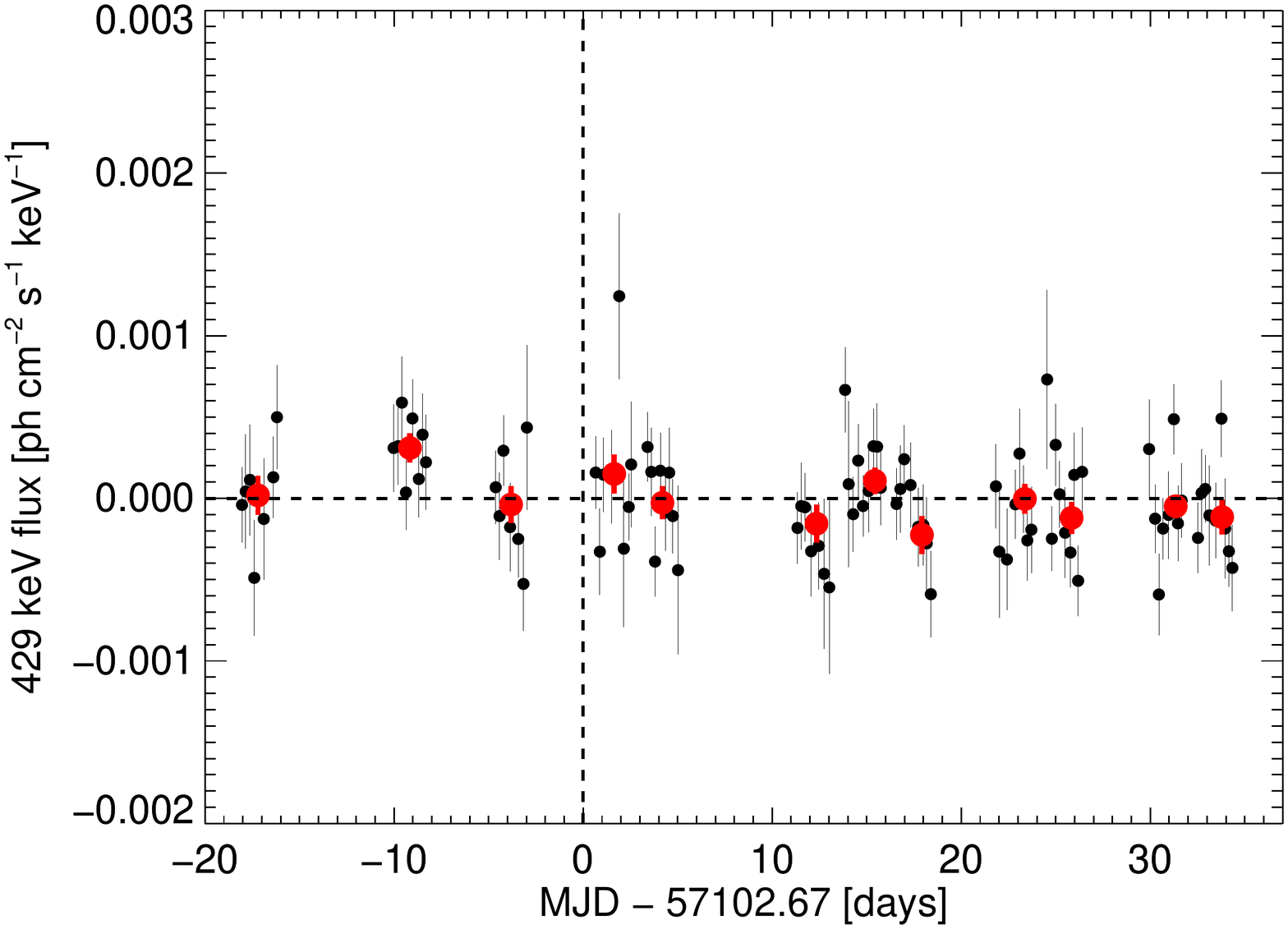}}\\
	  \subfloat[478~keV line. \label{fig:lc_478}]{\includegraphics[width=0.363\textwidth,trim=1.4cm 1.7cm 2.3cm 2.3cm,clip=true]{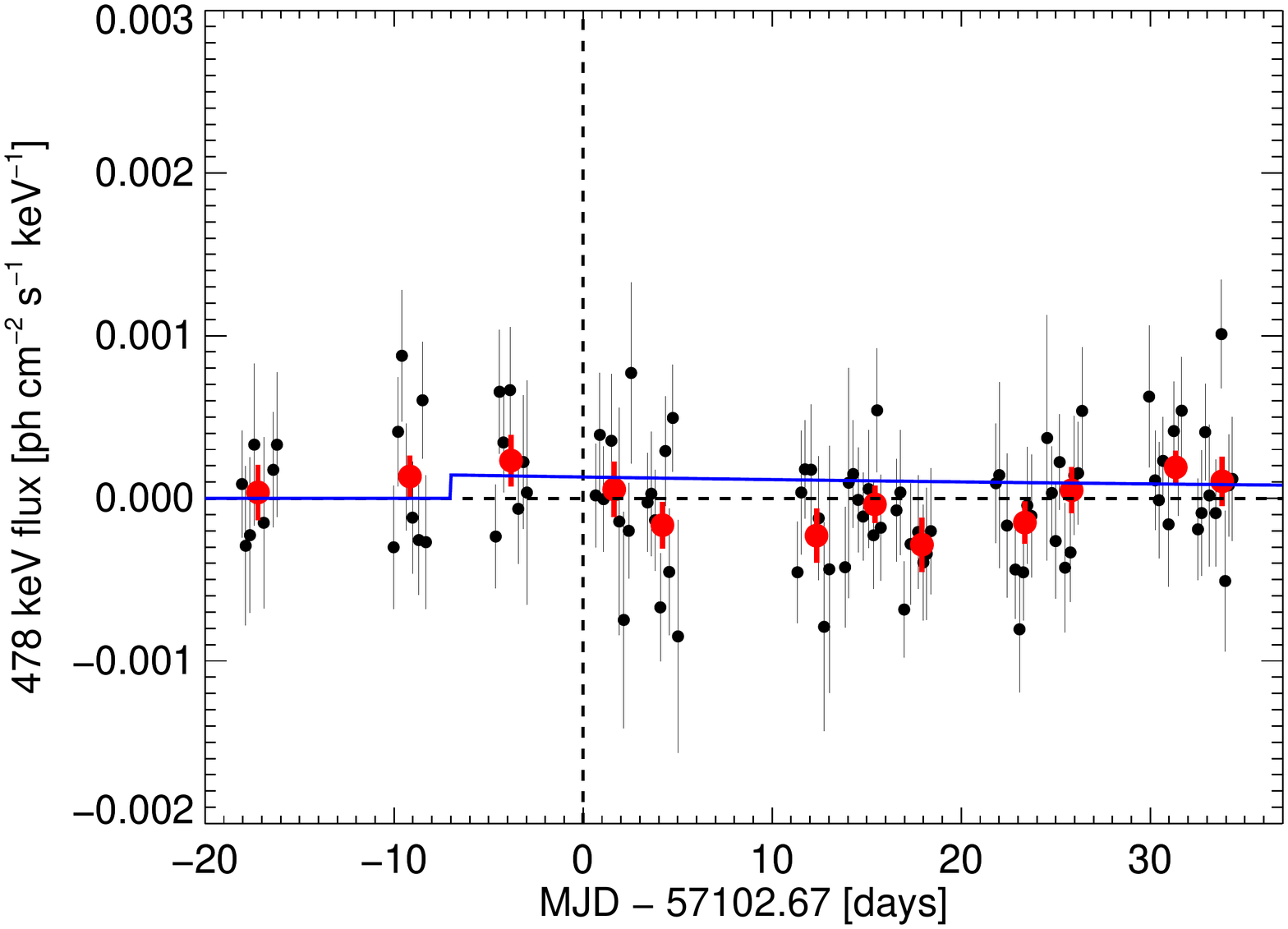}}\\
		\subfloat[511~keV line. \label{fig:lc_511}]{\includegraphics[width=0.363\textwidth,trim=0.5cm 1.6cm 3.0cm 2.9cm,clip=true]{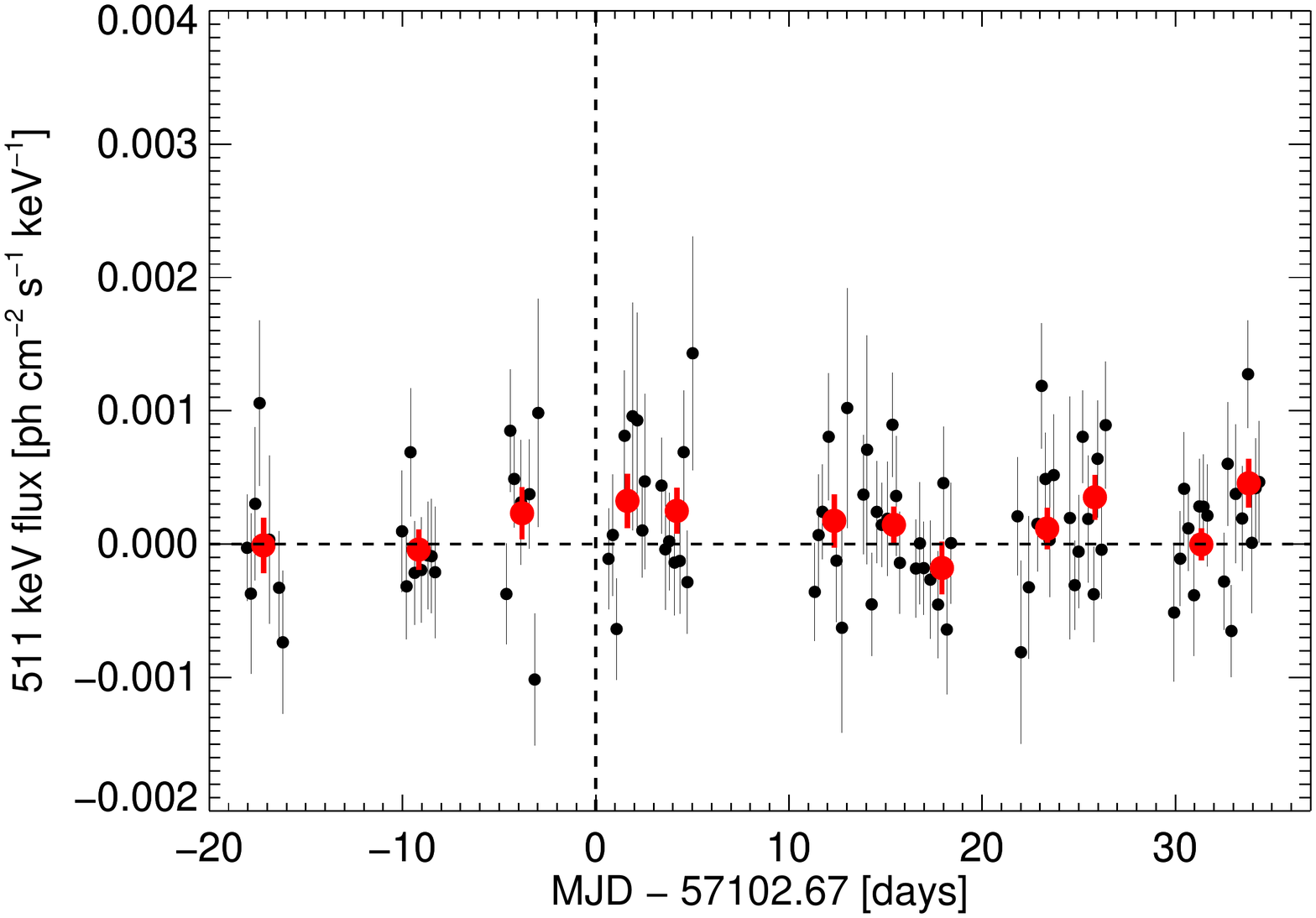}}\\
	  \caption{Light curves of the gamma-ray lines at 429~keV ($^3\mathrm{He}(\alpha,\gamma)^7\mathrm{Be^*}$, panel a), 478~keV ($^7\mathrm{Be}(EC)^7\mathrm{Li}$, panel b), and 511~keV ($\mrm{e^+ + e^- \rightarrow \gamma + \gamma}$), panel c. Shown are measured flux values in the energy bands 425-433~keV, 475-483~keV, and 506-516~keV, respectively, for 2-3 hours time bins (black dots), and $\approx 2$-day intervals (red dots), with respect to the optical maximum at $T_0 = \mathrm{MJD}~57102.67$ (vertical dashed line). The zero flux levels are indicated by a horizontal dashed line. No significant excess ($>3\sigma$) is found during the observations. In order to derive an upper limit on the flux from the decay of $^7\mrm{Be}$, an exponential radioactive decay law has been fitted to the data points. As the true time of explosive burning of V5668 is not known, a canonical time offset of $\Delta t = -7~\mrm{d}$ is used. The $3\sigma$ upper limit on this decay-law is shown as solid blue line in panel b). An upper limit on the annihilation flux at that time is not possible as INTEGRAL was pointed away from the source. See text for more details.}
		\label{fig:V5668_7Be_lc}
\end{figure}	

In Figs.~\ref{fig:lc_429}, \ref{fig:lc_478} and \ref{fig:lc_511}, the gamma-ray light curves of the 429, 478 and 511~keV lines are shown, respectively. There is no significant excess in the gamma-ray light curves from $^7\mrm{Be}$ or positron annihilation from the position of V5668. Around day -10, the mean 429~keV line flux has a significance of 3.6~$\sigma$ above the background. However, in the continuum band between 70 and 520~keV, no signal is detected ($\stackrel{\mrm{3\sigma}}{<}0.018~\mrm{ph~cm^{-2}~s^{-1}}$), and we consider this to be a statistical fluctuation. During days -8 to -6, for example, INTEGRAL observed other parts of the sky, and V5668 was not in the field of view (see angular position of V5668 with respect to the SPI on-axis frame in Fig.~\ref{fig:novadirection}). Either the source was in the field of view of SPI and IBIS, or the backside of the veto-shields are exposed to the source. Fitting the radioactive decay, Eq.~(\ref{eq:radiodecaylaw}), to the data in Fig.~\ref{fig:lc_478} obtains an upper limit on the synthesised $^7\mrm{Be}$ mass of $M^{7\mrm{Be}}_{3\sigma} < 6.4 \times 10^{-9}\,(d/\mrm{kpc})^2~\mrm{M_{\odot}}$. Using the 1.6~kpc distance estimate as before, the mass is constrained to $M^{7\mrm{Be}}_{3\sigma} < 1.6 \times 10^{-8}~\mrm{M_{\odot}}$. Assuming again the mass estimate by \citet{Molaro2016_V5668}, V5668 must be further away than $d^{\mrm{7Be}}_{3\sigma} > 1.1~\mrm{kpc}$. These limits become more constraining, if the time-dependence of the expected signal (radioactive decay) is taken into account, i.e. $M^{7\mrm{Be}}_{3\sigma} < 4.8 \times 10^{-9}\,(d/\mrm{kpc})^2~\mrm{M_{\odot}}$ and $d^{\mrm{7Be}}_{3\sigma} > 1.2~\mrm{kpc}$.

\subsection{ACS analysis}\label{sec:acsanalysis}

We use the SPI-ACS to search for burst-like emission during three weeks before the optical maximum of the nova. SPI was not pointed to the direction of V5668 for most of this time, and the ACS has an omni-directional field of view to possibly detect such a feature. The nova model light-curve by \citet{Hernanz2014_nova} provides a template for the temporal evolution of the expected gamma-ray flash. Between 75 and 511~keV, the light-curve rises with a maximum at hour 1, sharply decreases exponentially to hour 4, and then fades away. We interpolate this coarsely sampled model onto our one-minute time binning of the (total) ACS rate, $R_{ACS}^{tot}(t)$. In particular, we perform a search of excess signals in the ACS rate, i.e. nova flash candidate events, using a maximum likelihood method. This requires a background model for the ACS counts at any time, $B(t)$, an intensity scaling parameter, $a_0$, for the flash model, $F(t)$, and a temporal variable, $t_0$, determining the flash time. During the three weeks before the optical maximum, we test a grid of 200 amplitudes (source intensities), equally spaced between 0.0 and 0.2 of the peak amplitude of $0.21~\mrm{ph~cm^{-2}~s^{-1}}$, and use 400 time bins, equally spaced inside each INTEGRAL revolution, i.e. approximately 7.5-minute steps. Due to irregular particle events when entering and exiting the radiation belts, we limit the search to times when the ACS rate shows a smooth, non-erratic behaviour. This typically cuts out a few hours after the belt exits. We analyse INTEGRAL revolutions 1513 to 1521, i.e. days -20 to +3 in this way. The total model, $M(t;a_0,t_0)$, that is tested on the specified grid through the ACS rate in each orbit is then

\begin{equation}
M(t;a_0,t_0) = B(t) + a_0 \times F(t-t_0)\mrm{.}
\label{eq:acsmodelfit}
\end{equation}

As a background model for the ACS, we use a median filter of one hour applied to the ACS rate itself, $B(t) = \mrm{median}(R_{ACS}^{tot}(t),\mrm{60~min})$. This smears out features on this and shorter time scales, which may then be captured by our nova model light curve.

In each revolution, we find several candidates by iteratively accepting the (next to) maximum likelihood value in the marginalised probability density function of $t_0$. Table~\ref{tab:cands} summarises the nova flash candidate events with a statistical significance of more than $3.9\sigma$ ($p<10^{-4}$). In the following, we perform additional analyses to possibly distinguish between solar (particle and photon) events, gamma-ray bursts (GRBs), the nova itself, or other X-ray transients.

\subsection{Distinguishing between event candidates}

\subsubsection{Temporal characteristics}
The temporal characteristic of a candidate is a first indicator of its origin. While all candidates have been identified by using a nova model light-curve, any emission above the background, captured by this short-duration model, will improve the fit, but is not necessarily attributable to a nova. In general, a sharp rise and fast decay feature might point to a nova origin. However, also solar flare events show this behaviour, but in this case, this is followed by irregular and strong particle flux which is also measured by the ACS. Very strong gamma-ray bursts on time scales of seconds to minutes will also be captured by our method, but can easily be identified as such by investigating the residuals of the fit, as their temporal profiles are more stochastic than our smooth nova model. We provide the duration of the candidate event, $\Delta T$, as well as the $\chi^2$-goodness-of-fit value of the nova model in Tab.~\ref{tab:cands}, and mark possible GRBs and non-nova-like events in the comments column.

\subsubsection{Directional information}\label{sec:integralresponse}
The ACS consists of 91 individual BGO blocks, arranged in a hexagonal structure surrounding SPI up to its mask. If an event originates from a particular direction (point source), the facing side of the ACS will record more counts than the averted side. This can be expressed by an anisotropy factor of the different ACS sub-units, which consist of rings at different positions with respect to the SPI camera, and with different BGO thicknesses, ranging from 16~mm at the top (upper collimator ring, UCR1) to 50~mm at the lower veto shield directly below SPI. Most of the sub-units alone are not suitable to perform such an anisotropy analysis because they are (partially) shadowed by the other main instrument on INTEGRAL, IBIS \citep{Ubertini2003_IBIS}. On the level of the Ge detector array, the path between two opposing BGO crystals is blocked by the camera itself, so that the information may be skewed. By performing a similar analysis for a major solar flare, \citet{Gros2004_solarflare} concluded that the UCR1 is the most sensitive sub-unit to infer coarse directional information. \citet{Gros2004_solarflare} defined the anisotropy parameter by $A = (R - L)/(R + L)$, where $R$ and $L$ are the rates of opposing UCR1 hemispheres, i.e. the total count rates of three BGO detectors each. This allows to identify azimuthal directions if the source direction (aspect) is perpendicular to the crystal.

\begin{figure*}%
\centering
\includegraphics[width=1.25\columnwidth]{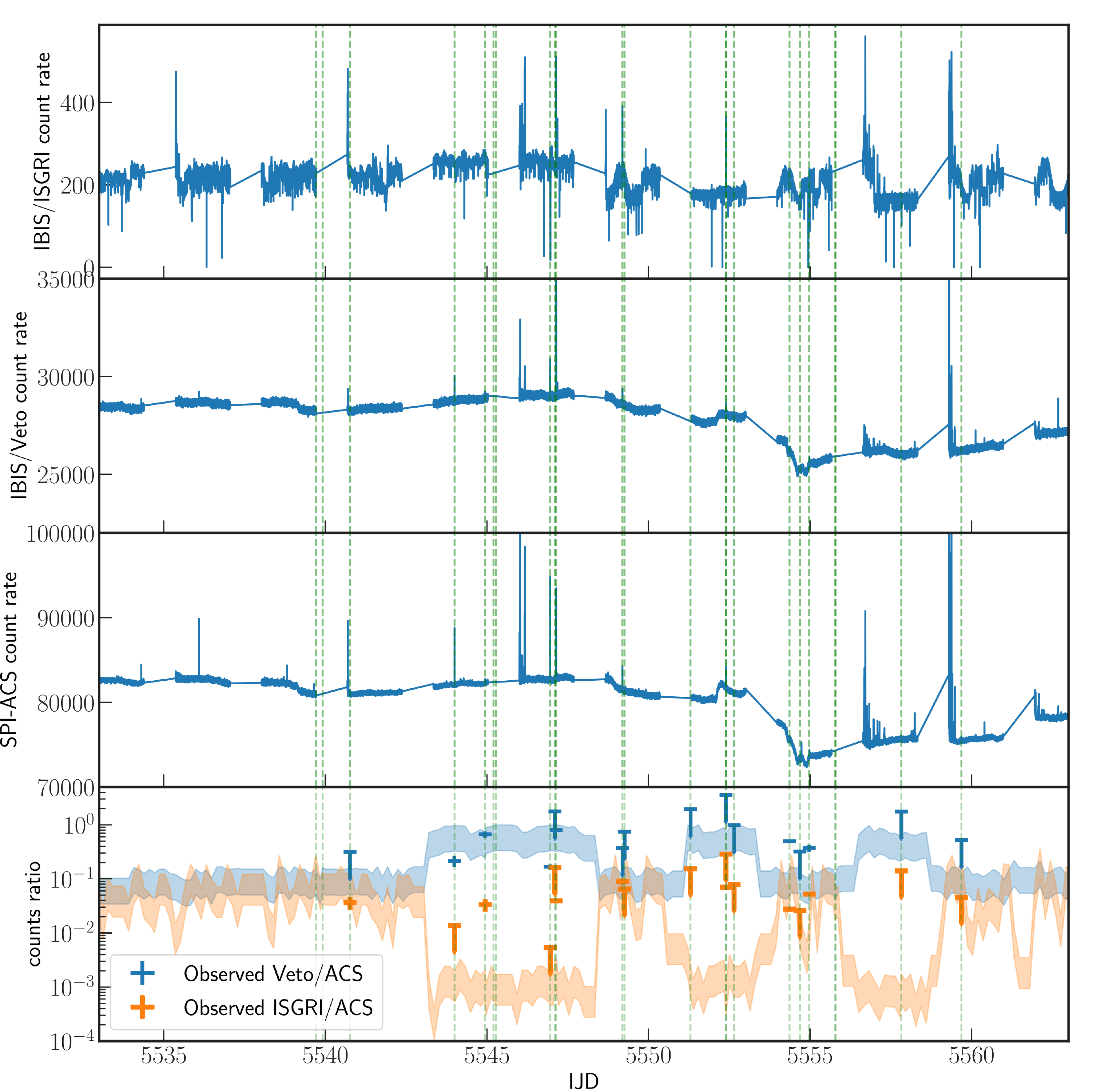}%
\caption{INTEGRAL response to the nova V5668 Sgr. The top three panels show the measured counts rates of different instruments aboard INTEGRAL in 100~s time bins (blue data points) From top to bottom, IBIS/ISGRI, IBIS/Veto, and SPI-ACS, respectively. The bottom panel shows the count ratios between the Veto and the ACS (shaded blue), and ISGRI and the ACS (shaded orange) if a source at the position of V5668 exposed INTEGRAL with a nova-like spectrum. The blue and red data points mark the actually measured ratios during the 23 event times (vertical dashed green lines). See text for details.}%
\label{fig:INTEGRALresponse}%
\end{figure*}

For general incidence angles to the veto shields, this anisotropy smears out, and the omni-directional INTEGRAL-response is required \citep[see e.g.][for the search of electromagnetic counterparts of gravitational wave]{Savchenko2017_LVT151012,Savchenko2017_GW170817}. We perform an analysis of the INTEGRAL veto-shields and ISGRI count rates. Depending on the source position (Fig.~\ref{fig:novadirection}) and the expected spectrum, the count rates vary accordingly. Comparing the ACS to Veto count ratio, and the ACS to ISGRI count ratio with the prediction from the source spectrum, a localisation is possible for strong sources, such as gamma-ray bursts. For the weak candidate events, we use this response to distinguish between a possible nova or another origin. In Fig.~\ref{fig:INTEGRALresponse}, the top three panels show the IBIS/ISGRI, IBIS/Veto, and SPI-ACS count rates, respectively. The bottom panel illustrates the expected count ratios from the direction of V5668 with estimated uncertainties (orange and blue bands), together with the actual measured count ratios for candidate events for which ISGRI data is available (cf. Fig.~\ref{fig:novadirection}, perigee passages). The spectral properties during the luminosity peak of the gamma-ray transient have been assumed to follow a power-law with a spectral index of $-1$. The angular dependence of the INTEGRAL all-sky response is mostly sensitive to spectral shapes between 50 and 300~keV, so that the model approximation remains valid although the spectrum, as estimated by \citet{Hernanz2014_nova}, may be more complex\footnote{Between 50 and 300~keV, the nova model by \citet{Hernanz2014_nova} indeed follows a power-law with index -1.}. For the purpose of localisation, it is sufficient to adopt a simple power-law spectrum. If the measured points coincide with this expectation, the event origin is likely to be from the direction of the nova, although may not be caused by the nova.

\subsubsection{Spectral hardness}
The distinction between solar events and X-ray transients can be augmented further by investigating the count rate in SPI during the time of a candidate event. While transient X-ray sources predominantly emit up to 500~keV photon energies, often with an exponential cut-off \citep[e.g.][]{Done2007_xrb}, solar flares can also show a strong increase in the high-energy continuum up to several MeV, and in addition de-excitation lines from $\mrm{^{16}O^*}$ (6.129~MeV), $\mrm{^{12}C^*}$ (4.438~MeV), and $\mrm{^{2}H^*}$ (2.223~MeV), for example \citep{Gros2004_solarflare,Kiener2006_solarflare}. Note however, that solar X-class flares which produce high-energy gamma-ray lines appear to be rare events \citep[$<10\%$,][]{Vestrand1999_SMMsolarflares}, but can be identified as such by SPI and its sub-systems. The SPI-ACS transparency is increased with increasing energy, so that we can utilise the information from the entire SPI instrument, i.e. ACS and Ge detectors. The measured residual Ge detector counts that pass the shield are thus a convolution of the ACS-transparency with the source emissivity. We define a hardness ratio between the energy bands 500-8000~keV and 20-500~keV, $HR_{500} = F_{500-8000} / F_{20-500}$, in one minute steps, to obtain another basis of decision for the nova flash candidates. In general, the hardness ratio $HR_{500}$ is a smoothly varying function with time and is typically around 0.4 in SPI raw data. This value can change over time, e.g. due to X-ray transients, showing a decreased $HR_{500}$ with respect to the average, whereas solar events increase the ratio. We analyse $HR_{500}$ during each candidate event, and compare the value to the averaged hardness ratio, integrated over one hour before and one hour after the event. We express the change in hardness in units of $\sigma$ in Tab.~\ref{tab:cands}. Positive deviations would indicate a solar origin, negative ones possibly X-ray transients.

\subsubsection{Event identification and discussion}
Based on the above criteria, we exclude nine of the 23 candidates by their temporal fitting residuals (GRBs, not nova-like in general), and five events are presumably from the direction of the Sun, based on an increased hardness ratio. Based on theoretical predictions \citep{Gomez-Gomar1998_novae} of such a nova flash 2 to 10 days before the optical maximum, we consider one event as too early ($t-T_0 = -429.8~\mrm{h} \approx -18~\mrm{d}$) to originate from V5668. Two other events may be considered too late, happening close or even after the V-band maximum. Several events coincide closely with GOES X-ray data, even though the hardness ratio is unchanged. Two additional events can be excluded in this way. The remaining three ACS features can be assigned a direction off the nova location, and do not show a decreased $HR_{500}$ which could be expected if a transient X-ray source was near. 

\begin{table*}[!ht]%
\centering
\caption{Nova candidate events, identified using the SPI-ACS rate and the nova model light curve by \citet{Hernanz2014_nova}. The columns contain the following values: The first column, $t-T_0$, shows the temporal difference between the optical maximum, $T_0$, and the event time in units of hours, $T_1$ and $T_2$ are the start and end times of the feature in units of MJD+57000, $\Delta T$ is the duration in units of s, $a_0$ is the fitted amplitude of the nova model, $\chi^2/\nu$ is the goodness-of-fit value, weighted by the number of degrees of freedom (reduced $\chi^2$), indicating whether a nova model is appropriate to express the light-curve\tablefootmark{b}, $\Delta L$ is the significance of the feature over the background in units of $\sigma$, and $HR_{500}$ is the hardness ratio difference, comparing the times before, during, and after the event in units of $\sigma$. The last column provides plausible source identifications, based on the other columns, coincidences with GOES X-ray data, or the INTEGRAL response. The horizontal lines indicate times for which the gamma-ray flash is expected to occur "most probably" \citep{Gomez-Gomar1998_novae}, i.e. between days 2 and 10 before the visual maximum.}\label{tab:cands}
\begin{tabular}{rrrrrrrrc}
\hline
$t-T_0$ & $T_1$ & $T_2$ & $\Delta T$ & $a_0$ & $\chi^2/\nu$ & $\Delta L$ $[\sigma]$ & $HR_{500}$ $[\sigma]$ & Comments \\
\hline
\hline
-455.2 & 83.713  & 83.715  & 181.3  & >0.2     & 234.6 & >1000 & -0.0 & GRB \\
-450.2 & 83.914  & 83.922  & 690.8  & 0.092(7) & 8.0   & 13.1  & +0.0 & not nova-like \\
-429.8 & 84.762  & 84.770  & 690.8  & 0.065(7) & 3.2   & 9.3   & -0.4 & too early \\
-352.5 & 87.993  & 87.994  & 25.7   & 0.049(8) & 22.3  & 6.1   & +2.2 & GRB \\
-329.7 & 88.943  & 88.950  & 604.5  & 0.050(7) & 3.3   & 7.1   & +2.9 & Sun / GOES \\
-322.0 & 89.205  & 89.223  & 1555.0 & 0.030(8) & 1.2   & 3.9   & +2.7 & Sun / GOES \\
-321.7 & 89.278  & 89.285  & 605.1  & 0.065(7) & 8.2   & 9.3   & -0.8 & not nova-like \\
-281.2 & 90.959  & 90.960  & 69.2   & 0.183(8) & 60.4  & 22.9  & +1.3 & GRB \\
-277.6 & 91.105  & 91.138  & 2851.0 & 0.054(7) & 85.9  & 7.7   & +3.0 & Sun / GOES \\
-277.0 & 91.140  & 91.142  & 129.2  & 0.155(7) & 120.0 & 22.1  & +1.5 & GRB \\
\hline
-227.4\tablefootmark{c} & 93.197  & 93.200  & 259.1  & 0.054(7) & 4.7   & 7.7   & +0.7 & not nova-like \\
-225.5\tablefootmark{c} & 93.260  & 93.305  & 3887.8 & 0.031(8) & 1.9   & 3.9   & +0.6 & GOES \\
-176.7\tablefootmark{c} & 95.300  & 95.340  & 3455.4 & 0.064(7) & 3.2   & 9.1   & +0.5 & false response \\
-150.8\tablefootmark{a} & 96.395  & 96.400  & 432.4  & <0.054   & 4.5 & <7.7  & +0.3 & weak \\
-150.8\tablefootmark{a} & 96.402  & 96.403  & 95.6   & <0.054   & 4.5 & <7.7  & +0.7 & GRB \\
-144.5\tablefootmark{c} & 96.650  & 96.661  & 950.5  & 0.036(8) & 1.9   & 4.5   & -0.0 & false response \\
-103.3\tablefootmark{c} & 98.362  & 98.405  & 3715.1 & 0.048(7) & 1.8   & 6.9   & +0.7 & false response \\
-95.6  & 98.687  & 98.714  & 2332.8 & 0.037(7) & 1.8   & 5.3   & +3.6 & Sun \\
-88.8\tablefootmark{c}  & 98.972  & 98.999  & 2332.8 & 0.034(7) & 2.2   & 4.9   & -1.7 & GOES \\
-69.1\tablefootmark{a}  & 99.781  & 99.782  & 86.4   & >0.2     & 18.2 & >1000 & -1.2 & GRB / GOES \\
-69.1\tablefootmark{a}  & 99.793  & 99.799  & 518.8  & <0.069   & 20.4 & <8.0  & +3.7 & Sun / weak \\
\hline
-20.3  & 101.825 & 101.866 & 3542.4 & 0.034(7) & 1.5 & 4.9   & +1.0 & too late \\
+24.3  & 103.686 & 103.699 & 1123.2 & 0.049(7) & 2.6 & 7.0   & -0.0 & after V-maximum \\
\hline
\end{tabular}
\tablefoot{
\tablefoottext{a}{Two events which are coincident within a few minutes are captured by the search method as only one. This happens e.g. when a strong GRB is preceding a solar particle event. Here, upper limits are given for fluxes and significances.}
\tablefoottext{b}{Because the nova light-curve model is not parametrised and thus very stiff, the resulting $\chi^2$-values are subject to extreme values and variations. We take this into account by considering also apparently bad $\chi^2$ values, up to $25\sigma$-deviations from the optimum. This declares reduced $\chi^2$-values below $\approx4$ as good, given the number of degrees of freedom of around 160.}
\tablefoottext{c}{Intriguing events discussed in the main text although formally excluded by specific diagnostics.}
}
\end{table*}

Even though our decision-making can exclude any of the candidates to be associated with V5668, the case for six of them remains intriguing. This may be either because the response is exactly met but the light-curve does not appear nova-like, or vice versa, or because an event is only close in time with GOES and not strictly coincident (marked (c) in Tab.~\ref{tab:cands}). These candidates would either be classified as "fast" ($T_0-5~\mrm{d}$ to $T_0-2~\mrm{d}$) or "moderately fast" ($T_0-10~\mrm{d}$ to $T_0-5~\mrm{d}$) novae, based on the gamma-ray flash occurrence time \citep{Gomez-Gomar1998_novae}. The measured peak fluxes are similar for all candidates, ranging between $6.5$ and $13.4 \times 10^{-3}~\mrm{ph~cm^{-2}~s^{-1}}$ in the energy regime of the SPI-ACS. These values are systematically uncertain by about 60\% because the true effective area at each individual event is not known. The statistical uncertainties range between 10 and 25\%. Taking all uncertainties into account, these events would correspond to luminosities $\approx 4$ to 50 times higher than the model assumption. This is a factor of 2-7 above the inferred values from \citet[][1.54~kpc]{Banerjee2016_v5668} or \citet[][1.6~kpc]{Jack2017_V5668}. This discrepancy might be due to our use of the nova model light-curve, its shape and its peak amplitude, being uncertain by about an order of magnitude \citep[e.g.][]{Hernanz1997_nova,Hernanz2005_novae,Hernanz2014_nova}. With increased information of nuclear cross sections and more elaborate modelling approaches, the estimates become more realistic, though still, large uncertainties can arise, e.g. when considering the initial conditions.

The ACS feature with the best-fitting temporal behaviour ($\chi^2/\nu = 1.8$) occurred 4.3 days (103.3 hours) before the optical maximum, with a significance of $6.9\sigma$ above the median ACS count rate. Its duration is at least 3700~s, and thereafter drowning into the background again. The hardness ratio is not significantly increased ($+0.7\sigma$), so that a solar or nova origin can neither be excluded nor suggested. The response function excludes this source by about $3\sigma$ to come from the direction of V5668. Based on only the hardness ratio, the feature around day 3.7 (hour 88.8) before the optical maximum ($\Delta HR_{500} = -1.7\sigma$) would suggest an X-ray source, though with low significance. During these two events, INTEGRAL was pointed towards V5668, and the source was in the partially coded field of view of IBIS and SPI. No hard X-ray / soft gamma-ray emission (70-520~keV) was detected during this time for both instruments (SPI: $<0.017~\mrm{ph~cm^{-2}~s^{-1}}$ at day -4.3, $<0.049~\mrm{ph~cm^{-2}~s^{-1}}$ at day -3.7; IBIS: $<0.003~\mrm{ph~cm^{-2}~s^{-1}}$; $3\sigma$ upper limits). The strongest signal in the ACS rate occurred 7.4 days (176.6 hours) before the visual maximum with $9.1\sigma$ above the background level. This event was preceded by another, but weaker ($<3.9\sigma$) flash-like signal. This is typical for solar particle events in which the gamma-ray emission is preceding the low-energy particles by a few minutes. Yet, the hardness ratio ($\Delta HR_{500} = +0.5\sigma$), which would be expected to be significantly increased in this case, is not constraining enough. The expected Veto-to-ACS ratio (response) is met, but the ISGRI-to-ACS ratio is off by one order of magnitude, so that the true origin is questionable. Three events might either be too short (950~s at day -6.0 (hour -144.5), 260~s at day -9.5 (hour -227.4)) or too weak ($(6.5\pm1.7) \times 10^{-3}~\mrm{ph~cm^{-2}~s^{-1}}$ at -9.4 (hour -225.5), viz. 5.7~kpc distance; SPI $3\sigma$ upper limit in the 70-520~keV band: $<0.017~\mrm{ph~cm^{-2}~s^{-1}}$) if the nova model was correct within a factor of three. However, especially the candidate at hour -227.4 before the optical maximum attracts attention, because it is the only event for which the response is exactly met. Although the gamma-ray light curve is not nova-like, this short peak may only be "the tip of an iceberg", so that most of the photons are either drowned in the background, do not escape, or are not produced, and the nova flash is only leaking for several minutes. Including the systematic uncertainties as described above, the flux for this event would be between $(11\pm7)\times10^{-3}~\mrm{ph~cm^{-2}~s^{-1}}$ (ACS, model-dependent) and $(200\pm150)\times10^{-3}~\mrm{ph~cm^{-2}~s^{-1}}$ (model-independent, for 260~s). 

In general, neither of the six signal excesses during the weeks before the optical outburst of V5668 can clearly be claimed to be due to the gamma-ray flash of explosive burning. On the other hand, the cases for origins other than the nova (or X-ray transients in general) are also only weak.

\section{Summary, discussion, and conclusions}\label{sec:conclusion}

\subsection{Nucleosynthesis ejecta}\label{sec:nucsysejecta}

We report an analysis of INTEGRAL gamma-ray observations of Nova Sgr 2015 No. 2 (V5668 Sgr). Novae are expected to produce significant amounts of $\mrm{^7Be}$. This large mass appeared to be detected for the first time for V5668 Sgr by observations of Be II lines in UV wavelengths. Although the $\mrm{^7Be}$ II doublet at $313.0583~\mrm{nm}$ and $313.1228~\mrm{nm}$, respectively, has only an isotopic shift of $\Delta \lambda = -0.161~\mrm{\AA{}}$ with respect to the $\mrm{^9Be}$ II doublet \citep{Yan2008_Beisotopeshift}, the high resolution spectra from HDT \citep[$R \approx 50000$, cf.][Subaru Telescope]{Tajitsu2016_V5668} or UVES \citep[$R \approx 100000$, cf.][VLT]{Molaro2016_V5668} can easily distinguish between the two isotopes for narrow components. There are also other lines from iron-peak elements, such as Cr II or Fe II, which could contaminate the $\mrm{^7Be}$ II measurements but which can also clearly be identified as such. \citet{Molaro2016_V5668} estimate the possible contamination of the equivalent width of $\mrm{^7Be}$ II absorption to about 3.5\%.

However, the absolute $\mrm{^7Be}$ mass estimates may be more uncertain, and given our limits, it is interesting to check how much larger it could be: \citet{Tajitsu2016_V5668} and \citet{Molaro2016_V5668} estimate the mass fraction of $\mrm{^7Be}$, following \citet{Tajitsu2015_novaBe7} and \citet{Spitzer1998_ISM}, by comparing the equivalent widths of a reference element to the $\mrm{^7Be}$ II doublet. Here, the authors used Ca, which is not a nova product, in particular the Ca II K line at 393.3~nm. The conversion of the equivalent widths to the respective column densities only works if the lines are not saturated and fully resolved. In addition, the covering factor of the nova shell should not be a strong function of wavelength, as otherwise, the Ca II K line could be intrinsically stronger or weaker absorbed. The compared isotopes must be in the same ionisation state to infer the column density ratios, which are then also the relative elemental abundance. This seems to be true since no doubly ionised nor neutral lines for Ca have been found \citep{Tajitsu2016_V5668,Molaro2016_V5668}. Once the abundance ratio $X(\mrm{^7Be})/X(\mrm{^{40}Ca})$ has been determined, an assumption on the Ca abundance delivers the $\mrm{^7Be}$ abundance in the nova ejecta. Here, the authors assume a solar Ca abundance, which might under-estimate the $\mrm{^7Be}$ abundance by $\approx 30\%$, due to the abundance gradient in the Milky Way \citep{Cescutti2007_abundancegradient}. The ejected $\mrm{^7Be}$ mass was then estimated by \citet{Molaro2016_V5668} by assuming a canonical ejected mass of $\approx 10^{-5}~\mrm{\Msol}$. In general, the ejected mass may range between $10^{-7}$ and $10^{-3}~\mrm{\Msol}$ for CO novae \citep{Bode2008_novae}. \citet{Banerjee2016_v5668} estimated a gas ejecta mass of $2.7$-$5.4\times 10^{-5}~\mrm{\Msol}$, based on a canonical gas-to-dust ratio between 100 and 200, and their measured dust mass of $2.7\times 10^{-7}~\mrm{\Msol}$. The authors assumed a distance to V5668 of 2~kpc in their calculations, so that the dust mass and hence the gas mass, normalised to our 1.6~kpc assumption, may be about 40\% smaller. Altogether, the total ejected mass may be a factor of few (2-5) larger than canonically expected. This would then also lead to an increase in the ejected $\mrm{^7Be}$ mass to a few $10^{-8}~\mrm{\Msol}$. If the $\mrm{^7Be}$ mass indeed was $2 \times 10^{-8}~\mrm{\Msol}$, this would be in tension with our derived upper limits on the mass if the distance of 1.6~kpc was correct.

With a half-life time of $\approx 53~\mrm{d}$, the radio-isotope $\mrm{^7Be}$ is decaying via electron capture to an excited state of $\mrm{^7Li}$, which de-excites by the emission of a gamma-ray photon at 478~keV. Using the spectrometer SPI aboard INTEGRAL, we searched for $\mrm{^7Be}$-line emission during the observations of V5668, which covered several weeks around the nova's optical maximum. From high spectral resolution as well as temporal (light-curve) analysis, we found no significant excess in the energy region of interest. We provide $3\sigma$ upper limits on the 478~keV line flux of $8.2 \times 10^{-5}~\mrm{ph~cm^{-2}~s^{-1}}$, which can be converted to an upper limit on the ejected $\mrm{^7Be}$ mass of $M^{7\mrm{Be}}_{3\sigma} < 1.6 \times 10^{-8}~\mrm{M_{\odot}}$. This, however, is based on uncertain distance estimates of 1.6~kpc. Assuming an ejected mass as derived by \citet{Molaro2016_V5668}, we can constrain the distance to V5668 to be further away than $d^{\mrm{7Be}}_{3\sigma} > 1.1~\mrm{kpc}$. Considering the detection of high-energy gamma-rays in the GeV range for V5668 Sgr, the fluxes around the 478 and 1275~keV lines may also have an underlying continuum from shock-accelerated particles. The estimated flux at both lines would be of the order $10^{-8}$-$10^{-7}~\mrm{ph~cm^{-2}~s^{-1}}$, following the description of \citet{Metzger2015_novashocks}. Even though this is below the sensitivity limit of SPI, and would only contribute less than 1\% to our upper limits, this effect is already accounted for in our derivation. Because approximated (Gaussian) line shapes and not only the flux values themselves are used, the line flux limits do not depend on the continuum below.

Nova Sgr 2015 No. 2 was identified to be a CO nova, and thus no to little $^{22}\mrm{Na}$ is expected to be produced and ejected. A gamma-ray line at 1275~keV would reveal the presence of $^{22}\mrm{Na}$, which is not seen in our analysis ($F^{22\mrm{Na}}_{3\sigma}<7.6 \times 10^{-5}~\mrm{ph~cm^{-2}~s^{-1}}$).

\subsection{Burst-like emission}\label{sec:burst_like}

Explosive nucleosynthesis in novae is accompanied by burst-like gamma-ray emission from short-lived isotopes. The $\beta^+$-decay of these isotopes is expected to be followed by positron annihilation in the nova cloud, leading to a strong annihilation line at 511~keV and continuum down to $\approx 20$-$30$~keV, depending on the conditions in the nova. Although this signal would be expected to be of the order of $0.1~\mrm{ph~cm^{-2}~s^{-1}}$ at 1~kpc in the 70-520~keV band, i.e. measurable in the SPI-ACS, it has never been observed, because it is expected to occur about one week before the optical maximum of the nova. Hence it may only be seen by chance or by a retrospective analysis of large data sets. The time of this gamma-ray flash is also uncertain, and may range and vary between 2 to 10 days before the optical maximum of a nova. SPI was not pointed to V5668 during the interesting time of the gamma-ray flash. But the INTEGRAL satellite with its main instruments and veto-shields has an almost omni-directional response. Therefore, we performed a search in the SPI-ACS data, using a nova light-curve model. Our search found 23 candidate events with significances above $3.9\sigma$ of which we identify six to be possibly associated with V5668, based also on the directional INTEGRAL-response. However, all six excess signals lack strong evidence to really originate from the nova, and all but one would suggest distances of more than 3~kpc (see discussion about positrons below).

Based on temporal, spectral, and directional information from multiple instruments aboard INTEGRAL, we illustrated a way of searching for X-ray transient features in archival data. Our search is similar to GRB-analyses \citep[e.g.][]{Rau2005_GRB}, but is augmented further by the combined use of energy- and angular-responses of the INTEGRAL veto-systems and main instruments. Up to date of this work, the INTEGRAL archive comprises 15 years of data. The estimated galactic-wide nova rate is $50^{+31}_{-23}~\mrm{yr^{-1}}$, and the local nova rate may range between $0.1$ and $0.5~\mrm{kpc^{-3}~yr^{-1}}$ \citep{Shafter2017_novarate}, so that tens of novae could be expected to be hidden in the current 15 years of INTEGRAL data. A thorough retrospective search for X-ray transient features in the INTEGRAL satellite's veto-systems might reveal an entire family of unobserved/unrecognised sources.

While model calculations provide estimates of how much material is produced and ejected, the true conditions shortly after explosive burning are uncertain. The short gamma-ray flash is believed to originate from the injection of positrons from the $\beta^+$-decay of, predominantly, $^{13}\mrm{N}$ ($\tau = 14.4~\mrm{min}$) and $^{18}\mrm{F}$ ($\tau = 158.4~\mrm{min}$) into the expanding envelope, where they annihilate with electrons and produce the 511~keV annihilation line and a continuum below. At this time, the envelope must be transparent enough to have the gamma-rays escaping, as otherwise no emission would be seen, and the expanding nova cloud would heat up by the absorption of these gamma-rays. Although $^{13}\mrm{N}$ and $^{18}\mrm{F}$ are short-lived, it might also be possible that their decay positrons escape from the nova in large amounts. The positron escape fraction, $f_{esc}$, could be added as a free parameter, similar to supernovae \citep[e.g.][]{Milne1999_SNIa}, which would allow diagnostics in the galactic-wide positron puzzle:

The strongest persistent and diffuse soft gamma-ray signal is the 511~keV line from electron-positron annihilation, presumably in the interstellar medium of the Milky Way. The morphology of the emission and the origin of the positrons are probably decoupled, because 1) there are more sources to explain the amount of positrons than is actually seen, and 2) the positrons annihilate at rest, which involves a significant deceleration (i.e. kinetic energy loss) from relativistic energies to less than 1~keV. This requires several 100~pc propagation distances in typical interstellar medium conditions \citep[e.g.][]{Alexis2014_511ISM}. Which sources contribute to what extent is only known very roughly \citep[see][for a review and a global measurement-based discussion, respectively]{Prantzos2011_511,Siegert2017_PhD}. Based on the apparent non-detection of V5668, novae could add to the reservoir of "longer-lived"\footnote{Longer-lived here means no prompt annihilation in the nova itself, but annihilation 0.01-10~Myr later in the interstellar medium.} positrons in the Galaxy, if the escape is larger than predicted. Depending on the nova type, of the order $10^{-7}$-$10^{-3}~\mrm{\Msol}$ of material may be ejected \citep[e.g.][]{Jose1998_novae,Starrfield1998_V1974}. The mass fractions of the dominant positron-producers\footnote{This is the case for both, CO and ONe novae. In ONe novae, however, of the order of $10^{-8}~\mrm{\Msol}$ of $^{22}\mrm{Na}$ are produced additionally, which is also a $\beta^+$-decayer, and which in any case contributes to the galactic positron content, because $^{22}\mrm{Na}$ has a half-life time of 2.75 years, i.e. at times when the nova is fully transparent and the ejecta further away from the white dwarf.} $^{13}\mrm{N}$ and $^{18}\mrm{F}$ is of the order $10^{-3}$ and $10^{-4}$ \citep{Jose2001_novaegamma,Jose2003_novae}, respectively, so that $10^{-8}$-$10^{-7}~\mrm{\Msol}$ of $^{13}\mrm{N}$ and $10^{-9}$-$10^{-8}~\mrm{\Msol}$ of $^{18}\mrm{F}$ are created. The decay modes for both isotopes are nearly 100\% positron emission, so that in total $10^{48}$-$10^{49}$ positrons are created - per nova event. Considering the global nova rate, the average number of positrons created by the population of novae in the Milky Way is $(0.9$-$25.8)\times 10^{42} \times f_{esc}~\mrm{e^+~s^{-1}}$, where $f_{esc}$ may range between 0 and 1. For example, if all positrons escape, this would make novae the dominant positron producers in the Galaxy. On the one hand, according to \citet{Gomez-Gomar1998_novae}, a 100\% escape would be in strong tension with simulations. On the other hand, a 1-10\% escape, as could be suggested for V5668, would contribute to about 1\% of the total required positron production rate to explain the 511~keV emission in the Milky Way \citep{Siegert2016_511}.

In this work, we showed that INTEGRAL/SPI is capable to detect a broad (8~keV, FWHM) $^{7}\mrm{Be}$ line at 478~keV from classical novae up to a distance of $\approx800$~pc with $5\sigma$ significance for an observation time of 1~Ms, starting at the visual maximum of the nova. This is derived from tight upper limits on the expected $^{7}\mrm{Be}$ line flux at 478~keV from the nova V5668 Sgr. In addition, we show that retrospective searches in archival INTEGRAL data can return valuable information for studies of X-ray transients. During the ongoing INTEGRAL mission, at least one such nova event could be expected.

\begin{acknowledgements}
This research was supported by the German DFG cluster of excellence 'Origin and Structure of the Universe'. The INTEGRAL/SPI project has been completed under the responsibility and leadership of CNES; we are grateful to ASI, CEA, CNES, DLR, ESA, INTA, NASA and OSTC for support of this ESA space science mission. LD and MH acknowledge support from the Spanish MINECO grant \protect{ESP2017-82674-R} and FEDER funds. JJ acknowledges support from the Spanish MINECO through grant \protect{AYA2014-59084-P}, the E.U. FEDER funds, and the AGAUR/Generalitat de Catalunya grant \protect{SGR0038/2014}. SS acknowledges partial support from NSF, NASA, and HST grants to ASU. TS thanks Francesco Berlato for Fermi/LAT analysis of the candidate events.
\end{acknowledgements}

\bibliographystyle{aa} 

\end{document}